\documentstyle[a4,12pt,psfig]{article}
\newcommand{\al}{\alpha}
\newcommand{\bt}{\beta}

\newcommand{\be}{\begin{equation}}
\newcommand{\ee}{\end{equation}}
\newcommand{\ba}{\begin{eqnarray}}
\newcommand{\ea}{\end{eqnarray}}
\newcommand{\de}{\delta}
\newcommand{\dd}{\partial}
\newcommand{\ga}{\gamma}
\newcommand{\db}{\bar{\partial}}
\newcommand{\fb}{\bar{f}}
\newcommand{\tet}{e_z^{~~a}}
\newcommand{\tec}{e_{\bar{z}}^{~~a}}
\newcommand{\teo}{e_0^{~~a}}

\newcommand{\bz}{\bar{z}}
\newcommand{\ra}{\rightarrow}

\newcommand{\vs}{\vspace{0.2cm}}
\newcommand{\vsa}{\vspace{0.4cm}}
\newcommand{\vsb}{\vspace{0.8cm}}
\newcommand{\half}{\frac{1}{2}}
\newcommand{\eps}{\varepsilon}

\newcommand{\dr}{e^{~~a}_\mu}
\newcommand{\scon}{\omega^{~~a}_\mu}
\begin{document}
\begin{titlepage}
\begin{flushright}
hep-th/9511211 \\
THU-95/33\\
November 1995
\end{flushright}
\vsa
\begin{center}
{\large\bf Some approaches to 2+1-dimensional gravity \vs\\
            coupled to point-particles          \vsa\vsb\\}
           M.Welling\footnote{E-mail: welling@fys.ruu.nl}
           \vsa\vsb\\
   {\it Instituut voor Theoretische Fysica\\
     Rijksuniversiteit Utrecht\\
     Princetonplein 5\\
     P.O.\ Box 80006\\
     3508 TA Utrecht\\
     The Netherlands}\vsb\vsa\\
\end{center}
\begin{abstract}
\end{abstract}

\end{titlepage}
\section*{Introduction}
In 1963 Staruszkiewicz considered for the first time gravity in 2+1 dimensions
coupled to point-particles \cite{eerst}.
In 1984 the subject was reconsidered by Gott, Alpert, Giddings, Abbot, Kuchar,
Deser, Jackiw and 't Hooft \cite{beginarticle}. In these articles they found
solutions for the gravitational field around N static point-particles. They
also solved the case of one spinning particle located at the origin. The key
result was that locally these space-times are flat except at the particles
positions. This implies that the gravitational field contains no degrees of
freedom as can also be deduced from a simple counting argument:
We have 6 independent metric components minus 3 first-class constraints minus 3
gauge-fixing conditions, resulting in 0 degrees of freedom. The reason that
these spaces are not completely trivial is because of non-trivial boundary
conditions on the (flat) coordinates. For instance, in the case of a massive
point-particle sitting at the rest in the origin, we have to cut a wedge out of
space-time and identify opposite points of the wedge. So the range of the
angular coordinate $\varphi$ is: $\varphi\in [0,2\pi(1-4mG)]$ (m is the mass of
the particle). In general this identification  condition on the coordinates is
not a simple rotation but a Poincar\'e-transformation (see section 1). Not much
later Deser and Jackiw found solutions for gravity with a cosmological
constant. The case $\Lambda>0$ (de Sitter space) coupled to 2 static, antipodal
particles was solved \cite{cosmo}. In a geometrical approach 't Hooft solved
the N-particle case  ($\Lambda=0$), and proved that a Cauchy-formulation was
possible within which no closed timelike curves could occur \cite{tHooft}. A
different view on the problem was provided by Achucarro and Townsend and later
by Witten \cite{Witten}. They considered a Chern-Simons theory with a
gauge-field $A_\mu$ taking values in the Poincar\'e-algebra, and proved that
this theory is equivalent to 2+1 dimensional gravity. Later Grignani and
Nardelli invented a consistent way to couple point-particles to this gauge
field \cite{CS}. This Chern-Simons approach is closely related to the
description of gravity using Ashtekar variables \cite{looprep}. It is however
not known to me if people considered the coupling of point-particles (For a
review on loop-quantisation one should consult the lecture by Kirill Krasnov in
this volume). Finally I would like to mention Waelbroeck's approach
\cite{Waelbroeck} who consideres these Ashtekar variables on a lattice in order
to obtain a finite set of degrees of freedom. This is an exact description
because only the handles and particles are the true degrees of freedom.

A first step towards quantisation was made by Mazur, 't Hooft, Deser, Jackiw
and de Sousa Gerbert who studied the scattering of 2 quantum particles.
\cite{scattering1}. Later the calculation was also done in the Chern-Simons
approach \cite{scattering2}. Carlip considered the scattering of N particles
where he stressed the role of the braid-group in 2+1-gravity
\cite{scattering2}. All these approaches have their own way of quantising the
theory. The problem is however that not all these quantisations seem to be
equivalent as shown by Carlip \cite{6ways}. In a way this is disappointing, but
it reflects the fact that the problems encountered in quantising
3+1-dimensional gravity still survive the dimensional reduction to 2+1
dimensions. As the theory contains no gravitons, and therefore has a finite set
of degrees of freedom, some of the problems must be connected with the
covariance of the theory under coordinate transformations. The hope is of
course that we are able to solve these problems in the much simpler model of
2+1-gravity. The big challenge will be to formulate a consistent second
quantised theory of 2+1 -gravity and look into the problems of renormalisation.
This issue has not been adressed to my knowledge up to now. Finally I would
like to mention that there is a close relationship of 2+1-gravity with the
theory of topological defects in condensed matter physics \cite{defect}.

In this review that is based on a lecture given at the Kazan Summer School
1995, we discuss 2+1-gravity coupled to point particles. In section 1 we
shortly look at  the solutions found in \cite{beginarticle} and \cite{cosmo}.
In section 2 we give the essential ingredients of the polygon-approach of 't
Hooft. In section 3 a short introduction is given to the Chern-Simons
formulation of 2+1-gravity. Finally we will discuss reference \cite{Welling}
where a new gauge is introduced which
is very convenient for the description of particles. Almost nowhere I will go
into the issues of quantisation deeply, but will restrict myself to general
remarks. On the different ways to quantise the theory excellent reviews
exist\cite{Loll}. I am well aware that this survey is only a poor selection out
of a vast amount of papers that have appeared on the subject. I only restrict
myself to the issues adressed at the summer school with the exception of the
second part of section 1 where gravity with a cosmological constant is treated.

\section{The Geometrical Appraoch}
In this section we will give some simple exact solutions of the gravitational
field surrounding point-particles. Both the case $\Lambda=0$ and $\Lambda\neq
0$ will be treated.

In 2+1 dimensions the Einstein equations with a cosmological constant are given
by the same expression as in 3+1 dimensions:
\be
G^{\mu\nu}-\Lambda g^{\mu\nu}=8\pi G T^{\mu\nu}
\label{Einst}
\ee
A special feature of 2+1 dimensions is however that we can express the
curvature-tensor in the following way:
\be
R^{\mu\nu}_{~~~\al\bt}=\varepsilon^{\mu\nu\lambda}
\varepsilon_{\al\bt\sigma}G^{\sigma}_\lambda
\ee
Using (\ref{Einst}) this can be written as:
\be
R^{\mu\nu}_{~~~\al\bt}=8\pi G\varepsilon^{\mu\nu\lambda}
\varepsilon_{\al\bt\sigma}T^{\sigma}_\lambda+\Lambda(\de^\mu_\al
\de^\nu_\bt-\de^\nu_\al\de^\mu_\bt)
\ee
This implies that outside sources the curvature must be constant.
\be
R\equiv R^{\mu\nu}_{~~~\mu\nu}=6\Lambda
\ee
{}From this we deduce that the gravitational field has no local degrees of
freedom (there are no gravitational waves). Setting $\Lambda=0$ for the moment
and considering a pointlike particle at the origin we have:
\be
T^{00}=m\de^2(\vec{r})~~~~~~~~T^{0i}=0~~~~~~~~T^{ij}=0~~~~~~~i,j=1,2
\ee
By symmetry arguments we have $g_{0i}=0$. The nontrivial Einstein equations are
given by:
\ba
G^0_0&=&-\frac{1}{2} {^{(2)}R}=8\pi Gm\de^2(\vec{r})\label{G00}\\
G_{ij}&=&-\frac{1}{2N}(D_iD_j-\gamma_{ij}D^2)N=0\label{G0i}
\ea
Here $\gamma_{ij}$ is the intrinsic 2 dimensional metric and $D_i$ is the
covariant derivative defined with respect to that metric, $^{(2)}R$ is the 2
dimensional curvature scalar. Futhermore N is the well known lapse function
which equals in this static case $g_{00}=-N^2$.
Taking the trace of equation (\ref{G0i}) we notice that we have to solve
\be
D^2N=0~~~~D_iD_jN=0
\ee
which is solved by $N=$constant. We redefine $t$ so that $N=1$. In 2 dimensions
we can always choose coordinates so that $\gamma_{ij}=e^{\phi}\de_{ij}$. Using
this in equation (\ref{G00}) we find:
\be
\vec{\nabla}^2\phi=-16\pi Gm \de^2(\vec{r})
\ee
This is easily solved using $\vec{\nabla}^2\ln r=2\pi \de^2(\vec{r})$:
\be
e^\phi=Cr^{-8Gm}
\ee
Absorbing the constant C into r we finally have:
\be
ds^2=-dt^2+r^{-8Gm}(dr^2+r^2d\varphi^2)
\ee
Because we know that the curvature vanishes everywhere except at the particles
positions we can transform to local flat coordinates:
\ba
\rho&=&\frac{r^\bt}{\bt}~~~~~~\bt=1-4Gm\\
\theta&=&\bt\varphi \\
ds^2&=&-dt^2+d\rho^2+\rho^2d\theta^2
\ea
Although the situation looks trivial now we have to be carefull. The new
coordinate $\theta$ ranges from 0 to $2\pi \bt$. So there is a deficit angle in
space (see figure 1)
\begin{figure}[t]
\centerline{\psfig{figure=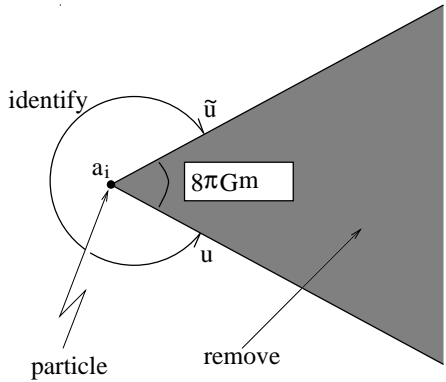,angle=-90,height=5cm}}
\caption{The wedge cut out of space-time}
\label{wedge1}
\end{figure}
The most important lesson we have to draw from this is that it is always
possible to transform to coordinates in which $g_{\mu\nu}=\eta_{\mu\nu}$, but
that there is a price to be paid. The price is that we have multivalued
coordinates (or coordinates with strange boundary conditions). This solution is
easily generalized to  N static particles. In this case it proves more
convenient to go to complex coordinates:
\be z=x+iy~~~~~~\bz=x-iy\ee
In these coordinates the line element for N static particles becomes:
\be
ds^2=-dt^2+\prod_{i=1}^{N} (z-a_i)(\bz-\bar{a}_i)^{-4Gm_i}dzd\bz
\ee
This is a multiconical space. Transforming to flat coordinates by using the
transformation:
\be
u=\int^z dz'\prod_{i=1}^N (z'-a_i)^{-4Gm_i}~~~~~~\bar{u}=c.c.
\ee
we construct a space from which we have to remove wedges emanating from every
particle. In this static case it is still unimportant in which direction we
choose the wedge. If however the particle moves, it is handy to choose the
wedge
behind the particle or in front of the particle. We will now argue why. In the
static case the wedge is characterised by an identification rule:
 \ba
\tilde{u}^a&=& R^a_{~b}u^b \label{transformationu}\\
R^a_{~b}&=& \left( \begin{array}{ccc}
          1&0&0\\
          0&\cos 2\pi\al&\sin 2\pi\al\\
          0&-\sin 2\pi\al&\cos 2\pi\al  \\
             \end{array} \right) \label{rotation}~~~~~~~\al=4Gm
\ea
Here $\tilde{u}^a$ is a point on one side of the wedge and $u^a$ the point on
the opposite side of the wedge. If the particle moves, the identification-rule
is still the same in its restframe. In the moving frame we therefore have:
\be
\tilde{u}^a = (BRB^{-1})^a_{~b}u^b\label{Lorentz}
\ee
Here $B^a_{~b}$ is a boost matrix with arbitrary rapidity $(\eta)$ and in an
arbitrary direction. In order to avoid a time jump we have to choose the wedge
symmetrically behind or before the particle. The effect is then that the wedge
becomes a bit larger. This is to be expected as the energy of the moving
particle is also larger.

The particle could also carry spin (classically). The energy momentum tensor
then looks like:
\be
T^{00}=m\de^2(\vec{r})~~~~~~~T^{0i}=\frac{1}{2}S
\varepsilon^{ij}\dd_j\de^2(\vec{r})~~~~~~~T^{ij}=0
\ee
Solving the Einstein equations is considerably more difficult ($g_{0i}\neq 0$)
and we will not repeat the derivation here. It can be found in
\cite{beginarticle}. Instead we will immediately give the result:
\be
ds^2=-(dt^2+2Adtd\varphi+A^2d\varphi^2)+(dr^2+r^2d\varphi^2)
\ee
Where $A=4GS$ is the spin of the particle. The important issue is again that we
can transform to the Minkowski line-element by the transformation:
\be
T=t+A\varphi
\ee
This has the strange consequence that we need a cut in space over which time
jumps by an amount $8\pi GS$. This also implies that close to the particle,
closed time-like curves are possible. Generalisations to N massive, spinning
particles exist \cite{Clement} but we will not treat that futher here. The
lesson is clear: Space-times with N moving, massive and spinning particles can
be constructed by cutting out wedges in space and define identifications over
the these wedges. Generaly, these identifications are an element of the
Poincar\'e group. As an example we treat the 2 particle case, with total
angular momentum J.
\begin{figure}[t]
\centerline{\psfig{figure=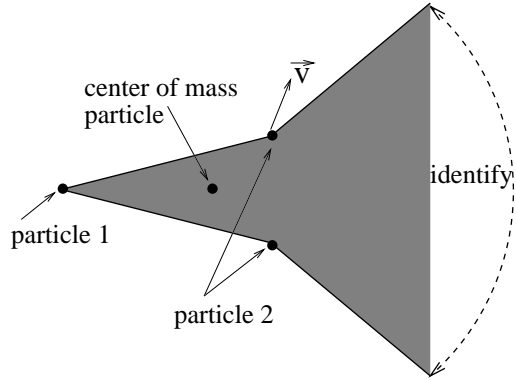,angle=-90,height=5cm}}
\caption{Wedge cut out around 2 particles.}
\label{2particles}
\end{figure}
We can always choose the particles on the x-axis. If we cut out the wedges as
in figure 2 we have the following identification:
\be
\tilde{u}=a_1+B_1R_1B_1^{-1}(a_2-a_1+B_2R_2B_2^{-1}(u-a_2))
\ee
where we didn't show the indices, and $a_i$ is the position of particle $i$.
Note that we didn't choose the wedge of the second particle at its tail,
introducing a time jump over the (total) wedge. The transformation is precisely
a Poincar\'e transformation:
\ba
\tilde{u}&=&\Lambda u+q\\
\Lambda&=&B_1R_1B_1^{-1}B_2R_2B_2^{-1}\\
q&=&B_1R_1B_1^{-1}(a_2-a_1)+a_1-B_1R_1B_1^{-1}B_2R_2B_2^{-1}a_2
\ea
If we write:
\be
\Lambda=B_{\bf com}R_{\bf com}B_{\bf com}^{-1}
\ee
this transformation really desribes a center of mass particle that is boosted
to a speed $v=\tanh\eta$. We still have the freedom to choose the overall
Lorentzframe so that we may take:
\be
\Lambda=R_{\bf com}
\ee
Futhermore we may place the c.o.m.-particle at the origin by demanding $q^i=0$.
Comparison with a spinning particle at rest in the origin suggests that the
total angular momentum is given by the time component of the translation vector
$q^a$:
\be
J=[B_{\bf
com}^{-1}q]^0=-[B_1R_1B_1^{-1}(a_2-a_1)]^0=-[B_2R_2B_2^{-1}(a_2-a_1)]^0
\ee
In the limit $G\ra 0$ we can indeed recover the special relativistic result.
In the next section we will also treat multiparticle solutions, but then we
will consequently put the wedges behind the particles in order to make a Cauchy
formulation possible (no time jumps). The total angular momentum expresses
itself then as a space-like translation over the total wedge.

Next we consider space-times with $\Lambda\neq 0$. The relevant reference is
\cite{cosmo}. Again we consider static configurations ($g_{0i}=0$) and choose
conformal coordinates on a time=constant slice: $\gamma_{ij}=e^{\phi}\de_{ij}$.
The Einstein equations for a static configuration of particles (without spin)
is:

\ba
&&\dd\db\phi+\frac{\Lambda}{2}e^{\phi}=-4\pi
GN^2\sum^{N}_{i=1}m_i\de^2(z-a_i)\label{liouville}\\
&&\dd\db N+\frac{\Lambda}{2}e^\phi N=0\\
&&\db V=\dd\bar{V}=0\\
&&V=\frac{1}{\Lambda}e^{-\phi}\dd
N~~~~~~~~~~\bar{V}=\frac{1}{\Lambda}e^{-\phi}\db N
\ea

\begin{figure}[t]
\centerline{\psfig{figure=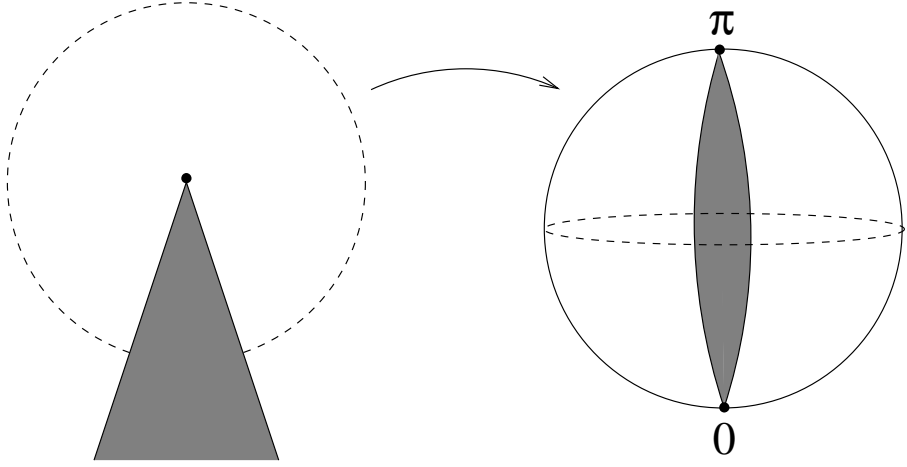,angle=-90,height=7cm}}
\caption{de Sitter space-time with a particle at the origin}
\label{embedding1}
\end{figure}
Here we used the notation: $\dd=\frac{\dd}{\dd z}$ and $\db=\frac{\dd}{\dd
\bz}$.
An additional condition is that initial static particles only remain static if
$\Gamma^i_{00}$ vanishes at the location of the source. One can show that this
implies: $NV=N\bar{V}=0$ at the source. Of course we must also insist that $N$
and $e^\phi$ are real and single-valued functions. Taking into account the
above considerations we can write down the following solutions for $\Lambda>0$:
\ba
&&e^\phi=\frac{4\dd f\db \bar{f}}{\Lambda(1+f\bar{f})^2}\\
&&N=\frac{1-f\fb}{1+f\fb}
\ea
where $f=f(z)$ is a holomorphic function and $\fb=\fb(\bz)$ is an
anti-holomorphic function. Also for $\Lambda<0$ we can check that we have the
solutions:
\ba
&&e^\phi=\frac{4\dd f\db \bar{f}}{|\Lambda|(1-f\bar{f})^2}\\
&&N=\frac{1+f\fb}{1-f\fb}
\ea
In order to reproduce delta-functions in equation (\ref{liouville}) we demand
that at $z=a_i$ we have the following singular behaviour for $\phi$:
\be
\phi\sim -4Gm_iN^2(a_i)\ln(z\bz) + \mbox{regular terms}
\ee
The case of 1 particle sitting in the origin is now easily solved by:
\be
f=z^{-\bt}~~~~\fb=\bz^{-\bt}~~~~~\bt=1-4Gm
\ee
We can check that $N$ is regular at the origin so that we can actually scale it
to 1 at $r=0$. We see  that in both cases the metric becomes pathological near
$r=1$. In the case $\Lambda<0$ this means that physical infinity is situated at
coordinate distance $r=1$. However in the case $\Lambda>0$ this is just an
artifact of the coordinate system and we may change to coordinates that cover
the whole sphere. In order to visualise these solutions it is convenient to use
the following embedding:
\ba Z&=&\frac{1-f\fb}{1+f\fb} ~~~~~~~~~~~\Lambda>0\\
U&=&\frac{2f}{1+f\fb}~~~~~~~~~~~~\bar{U}=c.c.\\
T&=&\frac{1+f\fb}{1-f\fb}~~~~~~~~~~\Lambda<0\\
U&=&\frac{2f}{1-f\fb}~~~~~~~~~~~~\bar{U}=c.c.
\ea
In the case $\Lambda>0$ this is just the stereographic projection, in the case
$\Lambda<0$ this is a projection from a hyperboloid to the plane. In the
$f$-plane the solution is pictured by cutting out a wedge emanating from the
particle's location $f=0$ (as in section 1 the angle variable of $f$ ranges
from 0 to $2\pi(1-4Gm)$). Mapping this on the sphere and the hyperboloid using
the above embedding results in the above pictures (figures 3,4).

\begin{figure}[t]
\centerline{\psfig{figure=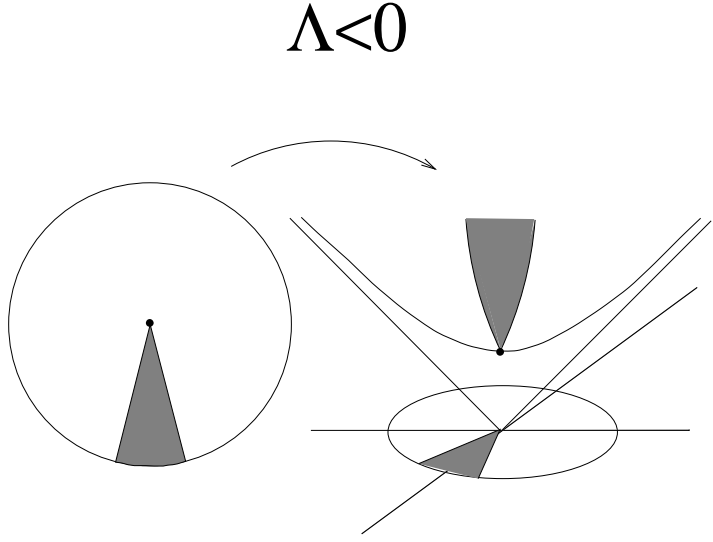,angle=-90,height=7cm}}
\caption{Anti de Sitter space-time with a particle at the origin}
\label{embedding2}
\end{figure}

In the case $\Lambda>0$ we see that the solution realy represents 2 antipodal
particles. In the second case we have found the Poincar\'e-disk
(Lobachevsky-space) from which a wedge is removed between 2 geodesics.
\footnote{Lobachevsky, who worked and lived most of his life in Kazan, was the
founder of non-Euclidean geometry.}
It is important to notice that in both cases a loop traversed around a particle
would result in a transformation:
\ba
&& f\ra e^{2\pi i\bt}f\\
&& \fb\ra e^{-2\pi i\bt}\fb
\ea
One can check that the expressions for $e^\phi$ and $N$ are invariant under
these transformations (as they should be by the requirement of
single-valuedness). In the case of de Sitter space one expects that static
configurations of  more than 2 particles do exist. It is however not known to
us if there exist explicit solutions in the literature.
\section{The Polygon Approach}

The relevant references are \cite{tHooft}.
The fact that the gravitational field has no degrees of freedom calls for an
approach in which only a finite set of degrees of freedom survive (the only
degrees of freedom of the theory are in fact the positions and momenta of the
particles). This is precisely what happens in what we call the Polygon approach
to 2+1-gravity coupled to point-particles. It was invented by 't Hooft to prove
that no closed time-like curves can occur in a closed universe. He proved that
the universe would crunch before the CTC could be finished. The basic idea
behind this method is to divide space up into rectilinear polygons with
vertices where 3 seams meet (see figure 5)
\begin{figure}[t]
\centerline{\psfig{figure=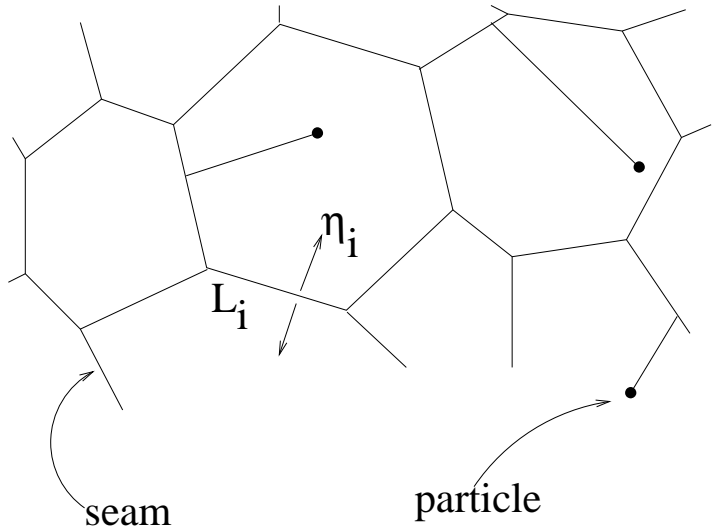,angle=-90,height=7cm}}
\caption{Space divided up in polygons.}
\end{figure}
\begin{figure}[t]
\centerline{\psfig{figure=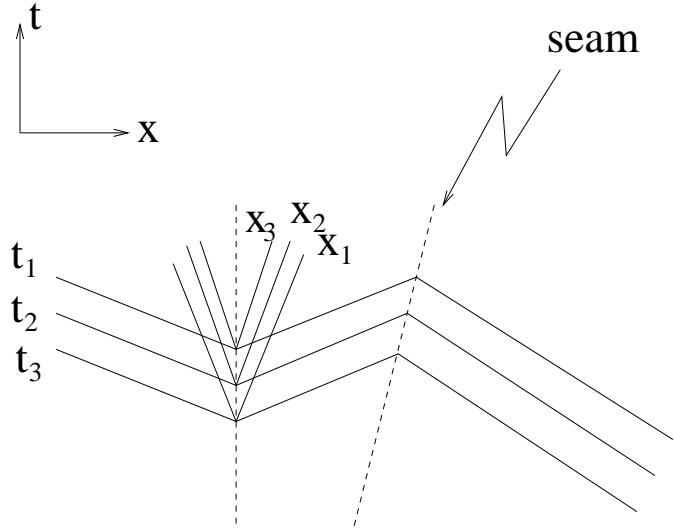,angle=-90,height=7cm}}
\caption{Polygons in the x-t axis}
\end{figure}

Just like in Regge-calculus, inside each polygon the space is flat and the
metric is simply Minkowskian. When we move over a seam and enter a new polygon
the coordinates we will change according to a Lorentz-transformation. Thus on
every polygon we choose different Lorentzframes. Futhermore we demand the
following 3 conditions:
\begin{itemize}
\item On every polygon there is a rest frame such that it is an equal time
surface.
\item At any time the total surface is a Cauchy surface and it is chosen to be
equal time everywhere.
\item The metric must be continuous at the seams.
\end{itemize}
Also particles are incorporated in this model. They sit at the end of a
1-vertex
(see figure 5). Using the above rules and our knowledge of 1 particle solutions
we can deduce the following rules:
\begin{enumerate}
\item The lengths of the seams are equal as considered from 2 adjacent
polygons.
\item The speed of the seams is always orthogonal to the seam and equal in
magnitude as considered from the 2 adjacent rest frames (see figure 7).
\begin{figure}[t]
\centerline{\psfig{figure=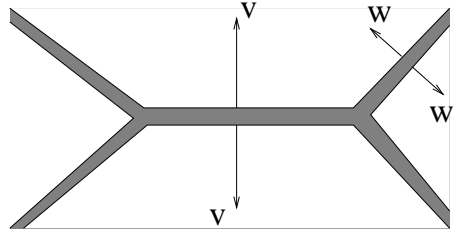,angle=-90,height=3cm}}
\caption{Moving seams}
\label{seams}
\end{figure}
\item The deficit angle is always in front or behind a particle (in order to
avoid time jumps; see figure 8).
\begin{figure}[t]
\centerline{\psfig{figure=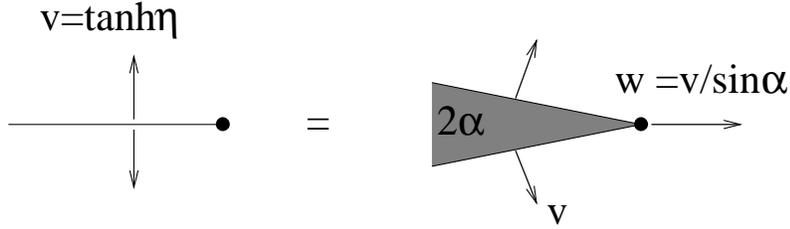,angle=-90,height=3cm}}
\caption{Wedge behind the particle}
\label{particle}
\end{figure}
\item At the 3-vertices (where no mass is present), there are relations between
the boost parameters $\eta_i$ and angles $\al_i$ (see figure 9).
\begin{figure}[t]
\centerline{\psfig{figure=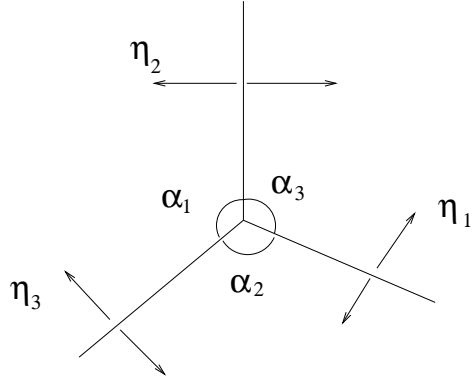,angle=-90,height=5cm}}
\caption{3-Vertex}
\label{3-vertex}
\end{figure}
These can be deduced from the fact that the 3 dimensional curvature must vanish
at the vertex (although the 2 dimensional curvature may be different from
zero).
Roughly speaking; 3 quantities determine 3 other quantities. One has to take
care however of inequalities such as the triangle inequality:
$|\eta_1|+|\eta_2|\geq|\eta_3|$.
\item As the system starts to evolve all kinds of transitions will take place
(actually 9 different kinds). We will only picture 2 examples in figure 10.
\begin{figure}[t]
\centerline{\psfig{figure=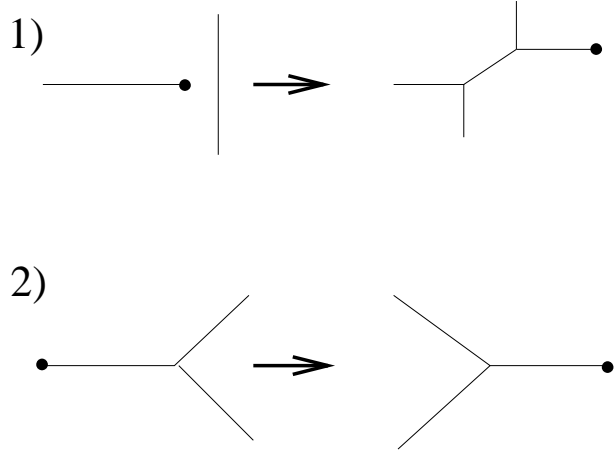,angle=-90,height=6cm}}
\caption{2 Possible transitions}
\label{transitions}
\end{figure}
The transition from one situation to the next is a completely determininistic
proces, i.e. all the new variables ($\eta_i,\al_i,L_i$) can be calculated from
the old ones. In this way we have a completely deterministic model, free of
pathologies with a finite set of degrees of freedom that can be used to study
2+1 dimensional gravity.
\end{enumerate}
It is also possible to write down a Hamiltonian formulation for this model. As
the total energy in 2+1 gravity is equal to the total deficit angle it is a
logical step to take the total sum of all deficits as the Hamiltonian of the
theory. Firstly we have a deficit angle of $H_P=2\al_P$ for a particle.
Secondly there is a possible deficit at a 3-vertex of
$H_V=2\pi-\al_{V}^1-\al_V^2-\al_V^3$ see figure 9). The total Hamiltonian is
now given by:
\be
H_{\bf tot}=\sum_i H_{P_i}+\sum_j H_{V_j}\\
\ee
As configuration variables we use the lengths of the seams: $L_i (\geq 0)$. We
already mentioned that is was possible to express the angles $\al_{V_j}^l$ in
terms of 3 neighbouring boostparameters $\eta_i$. Also the $\al_{P_i}$ can be
expressed in terms of the mass $m_i$ and boostparameter $\eta_i$ at his tail.
Moreover it is known how fast the lengths $L_i$ grow, i.e. $\frac{d}{dt}L_i$
can  also be expressed in terms of neighbouring $\eta_i$ (it is constant in
time).
We can use this information together with the equation of motion:
\be
\frac{d}{dt}L_i=\{H_{\bf tot},L_i\}
\ee
to derive that
\be
p_i=2\eta_i
\ee
The $L_i,2\eta_j$ are canonically conjugate phase space variables:
\be
\{L_i,2\eta_j\}=\de_{ij}
\ee
 The other equation of motion is trivial because the Hamiltonian
$H(\al_i(\eta_j))$ only depends on the momenta:
\be
\frac{d}{dt}p_i=\{H_{\bf tot},p_i\}=0
\ee
The simplicity is however a bit deceiving because we have to incorporate some
constraints in the model. The constraints could be expected since the number of
canonical variables exceeds the number of true degrees of freedom, which are
only connected with particles.
\footnote{The number of degrees of freedom for a closed (g=0) system is
calculated to be $4N-12$ where N is the number of particles.}
One can very easily understand these constraints. First of all the angles
inside the polygon must add up to $(m-2)\pi$ where m is the number of angles.
\be
C_1=\sum_{i=1}^m \al_i-(m-2)\pi=0\label{C1}
\ee
Secondly, the variables $L_i$, written as vectors, must add up to zero:
\be
\vec{C}_2=\sum_{i=1}^m\vec{L_i}=0
\ee
This can be reformulated in complex notation as:
\be
C_2=\sum_{k=1}^{m}L_ke^{i\theta_k}=0~~~~~~~~~~~
\theta_k=\sum_{j=1}^{k-1}(\pi-\al_j)\label{C2}
\ee
Futhermore we like to remind the reader that there are still the following
inequalities:
\ba
&&L_i\geq 0\\
&&|\eta_i|+|\eta_j|\geq|\eta_k|~~~~\mbox{and cyclic permutations}
\ea
which are hard to handle. The constraints (\ref{C1},\ref{C2}) are first class
and generate gauge transformations in the model. The first constraint evolves
one polygon in time. As a result the surrounding seams change their position.
The second constaint changes to a different Lorentz-frame inside the polygon.
Obviously, again all seams start to change. It can be calculated that these
changes are properly given by:
\be
\de L_i=\{L_i,C_j\}~~~~~~~~~\de p_i=\{p_i,C_j\}~~~j=1,2
\ee
One can also check that the constraint algebra closes properly. This is not a
trivial task at all and the algebra is highly nonlinear. With that we  mean
that it is not of the simple form $\{C_i,C_j\}=A^k_{ij}C_k$, but on the
righthand side for instance a $\sin(C_i)$ can appear. Of course the nontrivial
part of the evolution are the transitions that can take place and they have to
be dealt with separately. Now that we have formulated a Hamiltonian theory, the
way to quantisation seems to be open. One simply changes Poisson-Brackets by
commutators:
\be
\{.,.\}\ra-i\hbar[.,.]
\ee
Next one chooses for instance to work in the $L_i$ representation and writes
down a wave function:
\be
\Psi_D(L_1,...,L_N)
\ee
where D denotes its dependence on the diagram. First of all one needs to take
into account the transitions. They must be incorporated as boundary conditions
on the wave function. Needless to say that for a complicated diagram this is a
very difficult task. Also the constraints and inequalities should be taken into
account as constraints on the wavefunction to select the physical Hilbertspace.
Although some improvement can be gained by choosing clever coordinates it still
remains a very difficult task to complete this quantisation scheme. This will
be one of the motivations in section 4 to go to a single-valued coordinate
system. One can already deduce one important result in the quantum heory. As we
can only write down expressions for $\sin H$ and $\cos H$ we only have an
expression for the evolution-operator $e^{-iHt}$ and not for $H$ itself. This
implies that we cannot distinguish  between $e^{-iHt}$ and $e^{-i(H+2\pi n)t}$.
In order to avoid multi-valuedness we have to conclude that t is
integer-valued. So the model has discretised time!

\section{Gauge Theory of Gravity}
In this section we will briefly review some of the aspects of of the
Chern-Simons approach to gravity and the way it is coupled to point-particles.
The references are here (\cite{CS}). The main result is that the
Einstein-Hilbert action is equivalent to a Chern-Simons action where the gauge
field $A_\mu$ takes values in the Poincar\'e algebra ISO(2,1). If we write
$J^a=\half\eps^{abc}J_{bc}$ for the SO(2,1) generators and $P^a$ for the
generators of translations, the algebra is:
\be
[J^a,J^b]=\eps^{abc}J_c~~~~[J^a,P^b]=\eps^{abc}P_c~~~~~[P^a,P^b]=0
\ee
The gauge fields are then decomposed into:
\be
A_\mu=e^{~~a}_\mu P_a+\omega_\mu^{~~a}J_a\label{gaugefield}
\ee
Here $\dr$ (the dreibein or tetrad) is considered as the gauge field for
translations and $\scon=\half\eps^{abc}\omega_{\mu bc}$ (the spin connection)
as the gaugefield for Lorentztransformations. The usual Chern-Simons action is:
\be
I_{CS}=\int d^3x~\eps^{\mu\nu\rho}{\bf Tr}
(A_\mu\dd_\nu A_\rho+\frac{2}{3}A_\mu A_\nu A_\rho)
\ee
is invariant under the following gauge transformations:
\ba
\de A_\mu&=&-{\cal D}_\mu\Lambda(x)\\
&=&-\dd_\mu\Lambda-[A_\mu,\Lambda]
\ea
The gauge parameter $\Lambda$ can also be decomposed in terms of independent
gauge parameters for Lorentztransformations and translations:
\be
\Lambda=\rho^a(x)P_a+\kappa^a(x)J_a
\ee
The gauge transformations for the dreibein and the spinconnection are also
decomposed into local Lorentztransformations:
\ba
\de_1\dr&=&\eps^a_{~~bc}e^{~~b}_\mu\kappa^c\\
\de_1\scon&=&\dd_\mu\kappa^a+\eps^a_{~~bc}\omega_\mu^{~~b}\kappa^c
\ea
and local translations:
\ba
\de_2\dr&=&\dd_\mu\rho^a+\eps^a_{~~bc}\omega^{~~b}_\mu\rho^c\\
\de_2\scon&=&0
\ea
Witten has shown \cite{Witten} that these gauge transformations are equivalent
(if the equations of motion are satisfied) to the usual coordinate
transformations.
Substituting the expression (\ref{gaugefield}) into the C.S.-action and
performing a 2+1 split of space-time we can write down the following action:
\be
I_{\bf CS}=\int dt\int d^2x~~\eps^{ij}e^{~~a}_i\frac{d}{dt}
\omega_{aj}-\eta_{ab}e^{~~a}_{0}F_1^b[\omega]-\eta_{ab}\omega^{~~a}_0F_2^b[e]
\ee
One immediately reads from this action that $\eps^{ij}e^{~~a}_i$ is the
canonically conjugate variable to $\omega_{aj}$:
\be
\{\omega^{~~a}_i(x),\eps^{jk}e^{~~b}_k(y)\}=\eta^{ab}\de^j_i\de^2(x-y)
\ee
Futhermore, as there are no time derivatives of $e^{~~a}_0$ and
$\omega_0^{~~a}$, they act as Lagrange multipliers, imposing the constraints:
\ba
F_1^a[\omega]&=&\eps^{ij}R^a_{ij}~~~~~~~~\mbox{(curvature)}\\
F_2^a[e]&=&\eps^{ij}T^a_{ij}~~~~~~~~\mbox{(torsion)}
\ea
where:
\ba
R^a_{ij}&=&\dd_i\omega_j^{~~a}-\dd_j\omega_i^{~~a}+
\eps^a_{~~bc}\omega_i^{~~b}\omega_j^{~~c}\\
T^a_{ij}&=&\dd_i e_j^{~~a}-\dd_j
e_i^{~~a}+\eps^{a}_{~~bc}(\omega_i^{~~b}e_j^{~~c}+e_i^{~~b}\omega_j^{~~c})
\ea
These constraints are first class and it can be shown that they generate the
ISO(2,1) gauge transformations:
\be
\de G(e,\omega)=\{G,\rho(x)_aF_1^a[\omega]+\kappa_a(x)F_2^a[e]\}
\ee
Also the constraints obey the ISO(2,1) algebra. Finally, the usual expression
for the field-strength in a gauge theory is in this case translated to:
\be
F_{\mu\nu}=[{\cal D}_\mu,{\cal D}_\nu]=T^a_{\mu\nu}P_a+R^a_{\mu\nu}J_a~~~~(=0)
\ee
The next step is to couple the particles to this action. To keep the discussion
transparent we will couple {\em spinless} particles only, in the way proposed
by Grignani and Nardelli \cite{CS}:
\be
S_M=\sum_{i=1}^N\int d\tau~~\eta_{ab}p^a_i(\tau)D_\mu q^b_i(\tau)\dot{x}^\mu_i
+\lambda_i(p_i^2-m_i^2)
\ee
where $D_\mu$ is defined as the $ISO(2,1)$ invariant derivative:
\be
D_\mu q^a_i=\dd_\mu q^a_i+\dr+\eps^{abc}\omega_{\mu b}q_{ic}
\ee
This matter action is invariant under the gauge transformations:
\ba
\de_1 q^a&=&\eps^a_{~~bc}\kappa^bq^c\\
\de_1 p^a&=&\eps^a_{~~bc}\kappa^bp^c
\ea
\ba
\de_2 q^a&=&\rho^a\\
\de_2 p^a&=&0
\ea
In the paper of Grignani and Nardelli it is also stressed that the Poincar\'e
coordinate $q^a$ at this stage cannot be identified with a space-time
coordinate. Also the gaugefield $\dr$ is not the dreibein or soldering form and
the Poincar\'e torsion is not the space-time torsion. To make a connection with
a space-time interpretation we have to fix a gauge. For instance $q^a=0$ would
bring us back to the usual coupling of point-particles to gravity where $\dr$
can be interpreted as the dreibein. Another possibility would however be to
choose $q^a=\de^a_\mu x^\mu$.

The C.S.-theory explained above is a theory invariant under both
diffeomorphisms and Poincar\'e-gauge transformations (although they are not
independent). One way of quantising the theory is to construct a complete set
of gauge-invariant observables and use these as phase-space variables. In our
case we must find functionals of the gauge fields that are Poincar\'e-gauge
invariant and diffeomorphism invariant. These observables will then corresond
to Hermitian operators in the quantised theory. In the C.S.-theory these
observables can be found relatively easily. They are given by the Wilson-loops:
\be
W_R([\gamma])={\bf Tr}_R{\cal P}\exp[\oint_\gamma~A_\mu dx^\mu]
\ee
Here R denotes the representation used for the Poincar\'e generators. ${\cal
P}$ denotes path ordering, $\gamma$ is a spacelike loop and ${\bf Tr}$ denotes
that we have to take the trace. The argument of $W_R$ is denoted as $[\gamma]$.
This means that it is independent of the precise path of the loop but only
depends on the first homotopy class of loops on the punctured plane. This is a
difficult way of saying that all loops that can be deformed into each other
without moving over a puncture are considered as the same loop. The fact that
Wilson loops only depend on the homotopy class of loops is equivalent to the
statement that Wilson loops are invariant under diffeomorphisms (that are
continuously connected to the identity). These coordinate transformations
deform the path of the loop and the position of the particle in a continuous
and invertible way: $W_R(\gamma)\ra W_R(f({\gamma}))$. A diffeomorphism will
not move the path over a puncture so that
it remains in the same homotopy class. We will now argue why the Wilson loop is
invariant under these diffeomorphisms. It can be seen that the difference of
the deformed loop and the original loop is again a closed loop, not containing
any particles inside. Because the field strength inside this loop vanishes
everywhere it follows from the non-abelian Stokes theorem that this Wilson-loop
is actually the identity. Take for instance the simplified situation of figure
11:
\begin{figure}[t]
\centerline{\psfig{figure=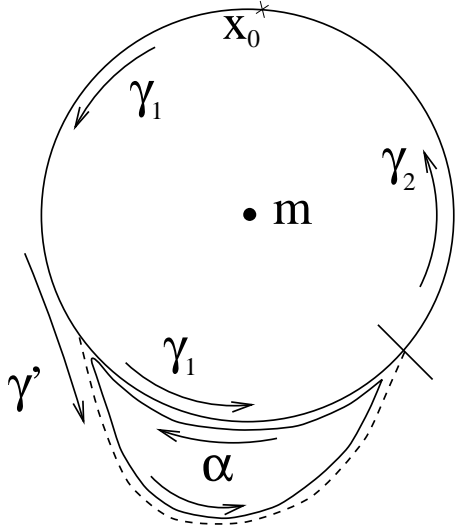,angle=-90,height=7cm}}
\caption{A Wilson-loop enclosing a particle}
\label{Wilson}
\end{figure}
\ba
W'_R(\gamma')&=& W_R(\ga_1)W_R(\al)W_R(\ga_2)\\
W_R(\al)&=&{\bf Tr}{\cal P}\exp[\int\int d^2x~~F_1^a[e]P_a+F^a_2[\omega]J_a]=I
\ea
It is also straightforward to show its invariance under gauge transformations
($g(x)$):
\ba
\tilde{W}_R(\gamma)&=&{\bf Tr}(g(x_0){\cal P}\exp[\oint
dx^\mu~A_\mu]g(x_0)^{-1})\\
&=& {\bf Tr}{\cal P}\exp[\oint dx^\mu~A_\mu]
\ea
Here we used the fact that the trace allows for cyclic permutation of matrices.
Martin \cite{Martin} calculated the explicit algebra of these Wilson-variables
and proposed to use this algebra as a starting point for quantisation. Of
course in the loop representation of Smolin and Rovelli \cite{looprep} it is
precisely these loops that act as fundamental variables in the theory. The wave
functionals then only depend on the homotopy class of loops. They proposed a
transformation from the gauge field representation to the loop representation
with precisely the Wilson-loop  as a kernel:
\be
\Psi([\gamma])=\int_{C/G}d\mu(A)~~W_R([\gamma])\Psi(A)
\ee
where $C/G$ is the space of all gauge inequivalent fields $A_\mu$ and $d\mu(A)$
is a measure on this space. More about the loop representation can be found in
Kirill Krasnov's lecture in this same volume.
\section{Gravity in 2+1 Dimensions as a Riemann-Hilbert Problem}
In this final section we will follow the lines of \cite{Welling} where we
define a new gauge that proves convenient in attacking the problem of solving
the gravitational field with point particle sources. We will work again in the
A.D.M. formalism and consider an open universe. The hope is that we will be
able to remove all redundant gravitational degrees of freedom from phase space
by solving them in terms of the particles positions and momenta. This process
is called reduction and works as follows: The total Lagrangian is schematically
written as:
\be
\frac{1}{16\pi G}\int d^3x~(\pi^{ij}\dd_t\gamma_{ij}-N_\mu H^\mu)+
\int dt~\sum_i p_{ia}\dd_tq^a_i-H_M+\frac{1}{16\pi G}\int d^3
x~\sqrt{-g}^{(2)}R
\ee
The first term is the Einstein-Hilbert action, the second is the particle
action and the third term is the surface term needed in the case of an open
universe \cite{ReggeTeitelboim}. Our gauge choice will remove the kinetic term
in E-H action. Solving the constraint equations at a time=constant slice will
remove all Hamiltonian terms. Inserting the solution of these constraint
equations into the boundary term will then generate the effective Hamiltonian.
After the reduction process we end up with:
\be
S_R=\int dt~p_{ia}\dd_t q_i^a+\frac{1}{16\pi G}\int dt\int
d^2x~\sqrt{-g\{g_{\mu\nu}(p,q,x)\}}^{(2)}R\{g_{\mu\nu}(p,q,x)\}
\ee
The surface term surface term is (after integration) an explicit function of p
and q. The fact that the gravitational field carries no degrees of freedom
makes it possible (in principle) to do this without losing any information. The
problem of N point particles was treated in section 2 with the use of flat
coordinates (the polygon-approach). We also mentioned that the unusual boundary
conditions on these coordinates made them multivalued. Consider for instance a
particle sitting at rest in the origin (see figure 12).
\begin{figure}[t]
\centerline{\psfig{figure=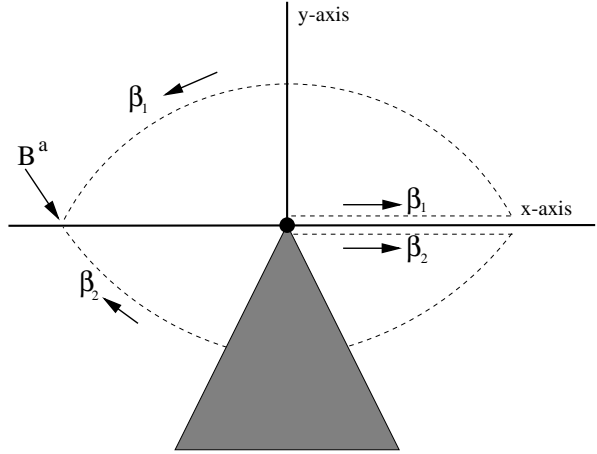,angle=-90,height=6cm}}
\caption{Particle sitting at rest in the origin}
\label{multivalued}
\end{figure}
An observer traversing via path $\bt_1$ to the point $B^a$ would give this
point different coordinates than an observer traversing path $\bt_2$. In the
following we will choose coordinates in such a way that we remove the boundary
conditions (i.e. we will have coordinates with the conventional ranges) but as
a price we will generate a non trivial interaction term.\footnote{This proces
is very much the same as in anyon-physics where also a multi-valued and
single-valued gauge exist} If we denote by $u^a$ the flat multi-valued
coordinates and by $x^\mu$ the curved, single-valued coordinates, the metric in
terms if the $x^\mu$ becomes:
\be
g_{\mu\nu}(x)=\frac{\dd}{\dd x^\mu}u^a(x)\frac{\dd}{\dd x^\nu}u^b(x)\eta_{ab}
\ee
First we like to choose a slicing condition. For that we view $u^a(x)$ as
embedding coordinates. So for every time $t$ we want to define a function
$u^a(x,y;t)$ that tells us how the $t=$constant surface is embedded in flat
3-dimensional Minkowski space-time. The condition will be that locally the area
of the surface must be maximal. This is of course always possible which assures
us that the gauge will be accessible. Actually it is precisely the Polyakov
action used in string theory that should be maximalised:
\be
S_P=\int d^2x~\sqrt{\ga}\gamma^{ij}\dd_i u^a\dd_j u^b\eta_{ab}
\ee
On the Cauchy surface we will choose conformal, complex coordinates, i.e.:
\be
\ga_{ij}=e^\phi\de_{ij}~~~~~~~z=x+iy~~~~~~~\bz=x-iy
\ee
The Polyakov action then reduces to:
\be
S_P=\int d^2z~\dd u^a\db u^b \eta_{ab}~~~~~\dd=\frac{\dd}{\dd
z}~~~~~~\db=\frac{\dd}{\dd\bz}
\ee
It is well known that the equations that follow from this action are:
\ba
\dd u^a\dd u^b\eta_{ab}&=&0\label{gauge1}\\
\db u^a\db u^b\eta_{ab}&=&0 \label{gauge2}\\
\dd\db u^a &=&0
\label{gauge3}
\ea
So $u^a$ must be a harmonic function: $u^a=u^a(z;t)+u^a(\bz;t)$. We will try to
solve the dreibein field $\dr=\dd_\mu u^a$ in the following. The conditions
(\ref{gauge1},\ref{gauge2},\ref{gauge3}) translate into the following
conditions for the dreibeins:
\ba
e^{~~a}_0\equiv \dd_t u^a &=& C^a(t)+\dd_t\int dz~\tet+\dd_t\int
d\bz~\tec\label{e0}\\
\tet\equiv \dd u^a &=&\tet(z)~~~\mbox{(holomorphic)}\label{ez}\\
\tec\equiv \db u^a &=&\tec(\bar{z})~~~\mbox{(anti-holomorphic)}\label{ebarz}\\
\tet e_z^{~~b}\eta_{ab}&=&0~~~~~~~~~~~\mbox{(null-vector)}\label{nullez}\\
\tec
e_{\bar{z}}^{~~b}\eta_{ab}&=&0~~~~~~~~~~~\mbox{(null-vector)}\label{nullebarz}
\ea
where $C^a(t)$ is some vector only depending on t.
Note that there is still conformal freedom: $z\ra f(z)$. These conditions in
turn can be translated into conditions on $\pi^{ij}$ and $\ga_{ij}$ (for the
definition of the canonical momentum $\pi^{ij}$ (see \cite{gravitation}):
\be
\ga_{ij}=e^\phi\de_{ij}~~~~~~~~~~~\pi\equiv\pi^i_i=0\label{york}
\ee
The advantage of these conditions is now evident. First of all, conditions
(\ref{york}) indeed imply that the kinetic term in the Einstein-Hilbert action
($\pi^{ij}\dd_t\ga_{ij}$) vanishes. To see that, one must split $\pi^{ij}$ and
$\ga_{ij}$ into a traceless part and the trace. As $\pi$ is conjugate to
${\bf Tr}\ga_{ij}=2e^\phi$ we find the result. Futhermore from a mathematical
point of view it is very convenient that $\tet$ is a holomorphic vector because
it allows us to  use the machinery of complex calculus.

Next we will try to reduce the problem of solving the gravitational field
around the moving particles to a mathematical problem, known as the
Riemann-Hilbert problem. First we mention that all the information for the
gravitational fields is really in $\tet$ and the asymptotic behaviour of the
gravitational field at infinity. If we know $\tet$ and the boundary conditions
at infinity we can in principle calculate $\tec$ and $\teo$. The asymptotic
behaviour can be studied by solving the gravitational field around a massive
spinning particle in our gauge. We expect that at $z\ra\infty$ this is the
asymptotic form of the multiparticle solution. The total mass of the universe
is the the mass of the particle and the total angular momentum is its spin (see
\cite{Welling}).
We have already argued that the flat coordinates $u^a$ are multivalued if we
traverse a loop around a particle. Specifically if we move around one particle
the result will only depend on the first homotopy class of the punctured plane
and not on the precise path chosen. The transformation will be:
\be
u^a\stackrel{[\ga_i]}{\ra}(B_iR_iB_i^{-1}\cdot u)^a +q_i^a
\label{transformationu2}
\ee
where $B_i$ is a boost-matrix, $R_i$ a rotation-matrix and $q_i^a$ is the
translation-vector.$[\ga_i]$ Denotes a loop in a certain class. We will look at
the transformation properties of $\tet=\dd u^a$:
\be
\tet\stackrel{[\gamma_i]}{\rightarrow}
(B_iR_iB^{-1}_i\cdot e_z)^a \label{transformatione}
\ee
The Riemann-Hilbert problem is now formulated as follows:

Given these set of monodromy transformations ($M_i=B_iR_iB_i^{-1}$) on a
Riemann surface, find the vector-functions $\tet$ that transform in this way
if we move around the puncture.

Because of the multivaluedness of $\tet$ we need cuts in the plane from the
$i^{th}$ puncture to infinity. $\tet$ Must now transform as it moves over the
cut. The familiar 1-dimensional example is of course $f(z)=z^\al$ where
$f\ra e^{2\pi i\al}f$ as we cross the cut. To treat the R-H problem it is more
convenient to use the 2 dimensional SU(1,1)  representation which is the
covering of SO(2,1). The R-H problem still remains the same only the monodromy
matrices $M_i$ now live in SU(1,1). From the spinor solution to the R-H problem
(denoted by $\zeta^\al$) we can then easily reconstruct the vector solution
$\tet$. It is also important to mention that the metric is single-valued (as it
should be).
\be
g_{\mu\nu}=\eta_{ab}\dr e^{~~b}_\nu\ra\eta_{ab}(M_i\cdot e_\mu)^a(M_i\cdot
e_\nu)^b=g_{\mu\nu}
\ee
Here we used the orthogonality property of SO(2,1) matrices:
\be
M^a_{~~b}M^b_{~~d}\eta_{ab}=\eta_{cd}
\ee
Another important remark is that the monodromy around the $z=\infty$ is
determined by the other monodromies around $z=a_i$:
\be
M_\infty^{-1}=M_N.....M_1\label{totmon}
\ee
This is due to the fact that a loop around infinity (on the Riemann-sphere) is
equivalent to a loop around all particles in the opposite direction.

There are 2 ways of studying the R-H problem and we will only briefly sketch
the methods here. In one approach one tries to write down a d'th order
differential equation with N+1 singularities (one of which is located at
infinity). d Is the dimension used for the representation (d=2 for SU(1,1) and
d=3 for SO(2,1)) and N is the number of particles involved. The singularities
must be of the special type called "regular singularities". It is always
possible by a global Lorentz transformation to go to a frame in which the
$i^{th}$ particle is not moving. This then implies that the monodromy around
that particle is given by a rotation. In this frame the spinor $\zeta^\al$ has
the following singular behaviour near the particle:
\be
\zeta^\al\stackrel{z\ra
a_i}{\sim}(z-a_i)^{\lambda^\al_i}[b^\al_i+c_i^\al(z-a_i)+...]
\label{localbehaviour}
\ee
where $\lambda^\al_i$ are called the local exponents. They must obey the
following rule:
\be
\sum_{\al=1}^d\sum_{k=1}^N \lambda^\al_k +\sum_{\al=1}^d
\lambda^\al_\infty=\frac{d(d-1)}{2}(N-1)
\label{somexponents}
\ee
In our case we will choose:
\be
\lambda^\al_i=(-\frac{\al_i}{2},\frac{\al_i}{2})~~~~~~~~~\al_i=4Gm_i
\ee
for the singularities in the finite part of the plane. The exponents at
infinity are then determined by relations (\ref{totmon},\ref{somexponents}). It
is important to notice that changing the local exponents by integers will not
change the monodromy. To solve the R-H problem uniquely we have to choose the
monodromy and the integer part of the exponents independently. One argument to
choose certain exponents could be that we demand that the singular behaviour
matches the behaviour found in pertubative calculations (small $Gm$). From
(\ref{localbehaviour}) we can easily deduce the the monodromy matrix:
\be
M_1=\left( \begin{array}{cc}
            e^{-i\pi\al_1} & 0\\
            0 & e^{i\pi\al_1}
            \end{array} \right)~~~~~~~~~\al_i=4Gm_i
\label{rot1}
\ee
The difficulty is of course that not all monodromies commute and can be brought
simultaneously to this diagonal form. For 2 particles however we can prove that
the most general second order linear differential equation with 2 singularities
and 1 at infinity can be written as a hypergeometric equation. It is long known
that the 2 independent solutions  of the hypergeometric equation transform into
a linear combination of them if we traverse a path around a singularity. By
matching this monodromy with the desired monodromy we can find a specific
solution to the R-H problem \cite{CS,Bellini,Welling}.
\footnote{Recently I received a note that Grignani and Nardelli actually
found this result first in the second reference of \cite{CS}.}

For more than 2 particles it is convenient to to consider a (equivalent)
linear, first order matrix differential equation of the form:
\be
\dd Y=\sum_{i=1}^{N}Y\frac{A_i}{z-a_i}
\label{dv}
\ee
$Y$ is the fundamental matrix of solutions to the differential equation. The
matrices $A_i$ are not depending on $z$ (but may depend on $a_i$ !). We demand
that the solutions to this equation must fullfill the following conditions:
\begin{enumerate}
\item[i)] $Y(z_o)=I$
\item[ii)] $Y(z)$ is holomorphic in $(C-\{a_1...a_N,\infty\})$
\item[iii)] $Y(z)$ has the following short distance behaviour near a singular
point:
\end{enumerate}
\be
Y(z)\stackrel{z\ra a_i}{\sim}(z-a_i)^{L_i}\hat{Y}_i(z)
\label{expY}
\ee
where $M_i=e^{2\pi i L_i}$ and $M_i$ are the monodromy matrices. It is also
important that the
matrix $\hat{Y}_i(z)$ is holomorphic and invertable at $z=a_i$. So:
\be
\hat{Y}_i(z)\stackrel{z\ra a_i}{\sim}(Y^0(a_i)+Y^1(a_i)(z-a_i)+....)
\label{hatY}
\ee
with $\det|Y^0(a_i)|\neq0$.
The R-H problem is now converted into finding the differential equation
(\ref{dv}) or equivalently to find the matrices $A_i$. This problem is in
principle solved in the mathematical literature \cite{Plemelj,Chudnovsky}.
Lappo-Danilevsky found an explicit solution to this problem in terms of a
series expansion in the $L_i$-matrices. He could prove convergence of this
series for small enough $L_i$. We will however not go into this technical
details of this solution. Miwa, Sato and Jimbo found a representation of the
$Y(z,a_i)$ in terms of a correlation function of conformal Dirac-spinors in a
free field conformal field theory. The particles are represented by
twistoperators and ensure the right monodromy properties if one moves around a
particle. We refer the interested reader to the literature
\cite{Moore,Sato,Blok,Welling}.
\section{Discussion}
In this survey we discussed some ways of treating 2+1-gravity coupled to
point-particles. We think this is an important issue because it is the starting
point for a quantised theory of gravity. One could for instance be interested
in defining a consistent S-matrix for quantised particles. First of all in- and
out-states should be defined with great care because the interaction is long
range. We expect however that consistent in- and outgoing states can be defined
because exact stationary scaling solutions exist (i.e. all particles move for
instance towards the center with a constant velocity proportional to the
distance from the center). It is however not clear if for some incoming
configurations the universe crunches and no outgoing states exist (or there may
for instance exist bound states of particles). Ultimately one would like to be
able to describe creation and annihilation of particles. Concluding we would
like to say that a lot of interesting work lies ahead.
\section{Acknowledgements}
I would like to thank the organising comittee of the Kazan State University for
their hospitality during my visit. Also I would like to thank the fellow
Phd-student for their friendship.

\end{document}